\documentclass[aps,prl,reprint,amsmath,amssymb,superscriptaddress,longbibliography,floatfix]{revtex4-1}
\usepackage{graphicx}
\usepackage{bm}   
\usepackage[hidelinks]{hyperref}
\usepackage{float}

\begin{document}

\title{Structured-Light Control of Goniopolar Thermoelectricity in NaSn\texorpdfstring{$_2$}{2}As\texorpdfstring{$_2$}{2}}

\author{Atoumane Ndiaye}
\email{ndiaya@rpi.edu}
\affiliation{Department of Materials Science and Engineering, Rensselaer Polytechnic Institute, 110 8th St., Troy, New York 12180, USA}
\author{Edwin Fohtung}
\email{fohtue@rpi.edu}
\affiliation{Department of Materials Science and Engineering, Rensselaer Polytechnic Institute, 110 8th St., Troy, New York 12180, USA}
\affiliation{Department of Physics, Applied Physics and Astronomy, Rensselaer Polytechnic Institute, 110 8th St., Troy, New York 12180, USA}
\affiliation{Center for Biotechnology and Interdisciplinary Studies, Rensselaer Polytechnic Institute,110 8th St., Troy, New York 12180, USA}

\date{\today}

\begin{abstract}
Goniopolar metals exhibit opposite thermoelectric polarities along different crystallographic directions, enabling zero-field transverse thermoelectricity but offering few means for external control. Here we show that a spatially structured Laguerre--Gaussian vector potential can programmably reconstruct the goniopolar phase space of NaSn$_2$As$_2$. First-principles-derived Wannier transport with bond-dependent Peierls coupling reveals two OAM-dependent spatial scaling laws: the radial response follows the Laguerre--Gaussian radius $r_{\max}\propto\sqrt{|\ell|}$, while calculations for $|\ell|=2$--5 yield a dominant angular harmonic $m_{\mathrm{dom}}=2|\ell|$, encoding the optical winding in a frequency-doubled thermoelectric response. Increasing $|\ell|$ simultaneously reconstructs pre-existing goniopolar windows, enhancing a representative window by approximately 17\% at $|\ell|=5$. By contrast, reversing $\ell$ at fixed polarization produces only a small correction that approximately interchanges upon polarization reversal. Energy-resolved transport reveals that the structured field redistributes in-plane and cross-plane electronic velocities, shifting the directional Seebeck-zero boundaries that define the goniopolar state. These results establish vortex position and OAM magnitude as programmable control coordinates for goniopolar thermoelectricity.
\end{abstract}

\maketitle

Goniopolarity is the unusual situation in which the effective carrier polarity depends on transport direction. In layered NaSn$_2$As$_2$, an open, concave Fermi surface produces electron-like in-plane and hole-like cross-plane response within the same electronic system \cite{He2019}; anisotropic scattering modifies the thermopower magnitude while preserving the central role of Fermi-surface geometry \cite{WangNarang2020}. Because $S_{ab}S_c<0$ generates a transverse thermoelectric response when the thermal gradient is rotated away from the principal axes, goniopolarity provides a zero-field route to transverse energy conversion \cite{UchidaHeremans2022,Adachi2025}. Recent work has broadened goniopolarity beyond NaSn$_2$As$_2$; a complementary challenge is whether the boundaries and spatial distribution of an existing goniopolar state can be controlled externally \cite{Hu2026}.

Structured light provides such a control coordinate because its transverse amplitude and phase carry spatial information unavailable to a plane wave \cite{Forbes2021,QuinteiroRosen2022}. Laguerre--Gaussian (LG) modes contain the azimuthal factor $e^{i\ell\phi}$ and carry orbital angular momentum (OAM) indexed by the integer $\ell$ \cite{Allen1992}. Importantly, structured light has now also been demonstrated experimentally as a control field for collective material order: twisted ultraviolet LG illumination of quasi-two-dimensional CsBiNb$_2$O$_7$ reconfigured ferroelectric polarization textures, ionic displacements, and strain fields resolved by operando coherent X-ray imaging and Raman spectroscopy \cite{Nazirkar2025}. This establishes an experimental materials-control precedent for using the spatial and topological degrees of freedom of light to reorganize an anisotropic condensed-matter state. The present work asks a distinct transport question: can that broader control paradigm be extended from real-space ferroic order to the momentum-space transport geometry underlying goniopolarity?

Here we demonstrate theoretically that structured light provides an
external control parameter for the goniopolar phase space of
NaSn$_2$As$_2$. Starting from a first-principles-derived Wannier
Hamiltonian, we couple the spatially varying LG vector potential
$\mathbf{A}_{\ell}(\mathbf{r})$ to the electronic hopping through a
bond-dependent Peierls phase
\cite{GeneralizedPeierls2022}. We find that the goniopolar-window
width $W_{\mathrm{GP}}$ develops an annular spatial modulation whose
radius follows $r_{\max}\propto\sqrt{|\ell|}$, while independent
calculations for $|\ell|=2$--5 reveal the angular scaling
$m_{\mathrm{dom}}=2|\ell|$. Increasing $|\ell|$ systematically
reconstructs multiple goniopolar windows, including an approximately
17\% enhancement of GP5 at $|\ell|=5$, whereas the difference between
opposite windings at fixed polarization remains comparatively small. A four-state
comparison that also reverses the polarization shows that this smaller
sign-dependent correction approximately follows the relative
OAM--polarization configuration rather than OAM sign alone.
Energy-resolved transport calculations
trace this behavior to a structured-field-induced redistribution of
in-plane and cross-plane electronic transport, which displaces the
Seebeck sign-change boundaries. These results establish structured
light as a spatially programmable control parameter for goniopolar
thermoelectricity and connect the spatial structure of an
electromagnetic field to the tunability of anisotropic thermoelectric
polarity.

\begin{figure}[t]
\centering
\includegraphics[width=1\columnwidth]{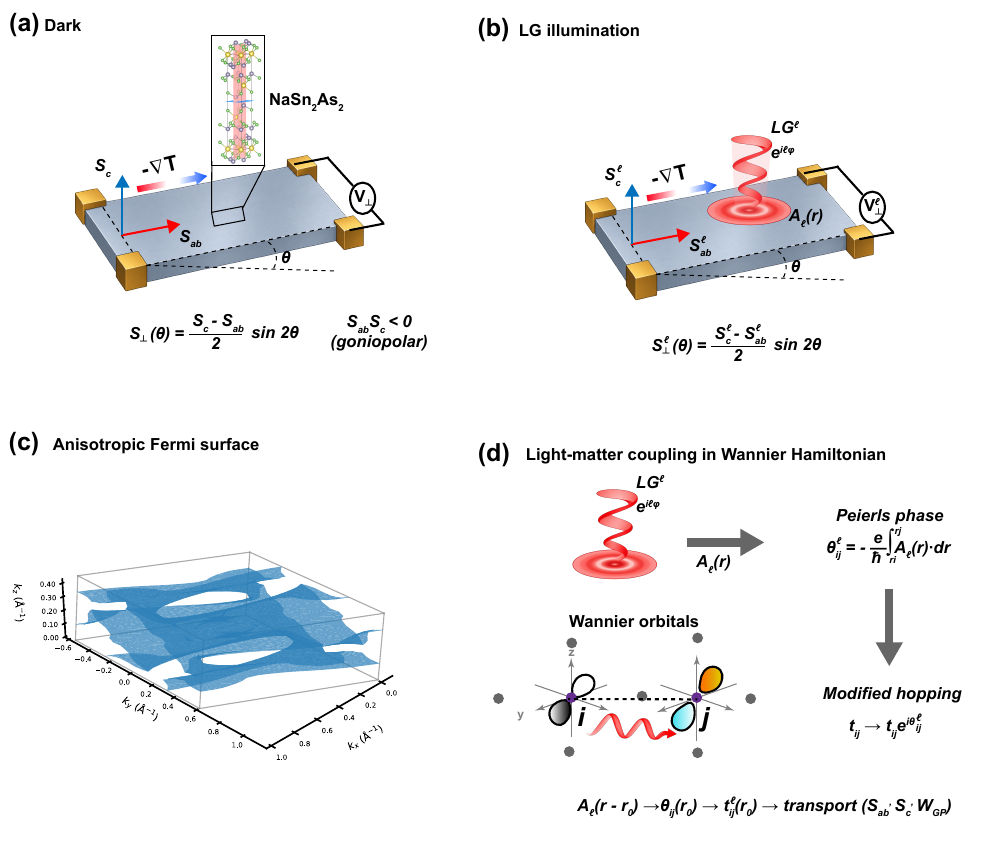}
\caption{
\textbf{Electronic origin and structured-light control of goniopolar thermoelectric transport in NaSn$_2$As$_2$.}
(a) Dark-state goniopolar geometry, with $S_{ab}S_c<0$. (b) LG structured field displaced relative to the crystal. (c) Anisotropic Fermi surface underlying direction-dependent transport. (d) Peierls coupling: $\mathbf A_\ell\rightarrow\theta_{ij}^{(\ell)}\rightarrow t_{ij}e^{i\theta_{ij}^{(\ell)}}$, which reconstructs $S_{ab}$, $S_c$, and $W_{\rm GP}$.
}
\label{fig:fig1}
\end{figure}

The electronic structure was obtained with PBEsol DFT and PAW pseudopotentials in \textsc{Quantum ESPRESSO}; the localized transport model was constructed with \textsc{Wannier90} and evaluated with semiclassical Boltzmann transport \cite{Giannozzi2017,Perdew2008,Blochl1994,Marzari2012,Pizzi2020,Madsen2018}. Fully relativistic calculations characterize the equilibrium fermiology shown in Fig.~\ref{fig:fig1}(c), whereas all dark/structured-field transport comparisons use the same scalar-relativistic Wannier Hamiltonian. This internal matching isolates the field-induced modulation, but an SOC transport benchmark remains desirable for its quantitative magnitude; detailed numerical parameters are given in the Supplemental Material.

We use the chemical-potential shift $\Delta\mu=\mu-\mu_0(T)$ and identify a goniopolar (GP) interval whenever $S_{ab}(\mu)S_c(\mu)<0$. For a pocket bounded by $\mu_n^-$ and $\mu_n^+$, $W_n=\mu_n^+-\mu_n^-$.

To investigate structured-field control, we consider a prescribed
$p=0$ Laguerre--Gaussian (LG) vector-potential texture carrying orbital
angular momentum (OAM) with winding index $\ell$. Its spatial envelope
is proportional to
\begin{equation}
\psi_{\ell}(\rho,\phi)
\propto
\left(\frac{\sqrt{2}\rho}{w_0}\right)^{|\ell|}
\exp\left(-\frac{\rho^2}{w_0^2}\right)
e^{i\ell\phi},
\label{eq:lg}
\end{equation}
where $w_0$ is the characteristic transverse scale. For
$|\ell|>0$, the radial envelope reaches its maximum at
\begin{equation}
r_{\max}=w_0\sqrt{\frac{|\ell|}{2}}.
\label{eq:rmax}
\end{equation}
Unless otherwise stated, we use $w_0=20$~\AA{},
$A_0=5\times10^{-9}$~T\,m, and transverse helicity $\sigma=+1$.

The structured field is coupled to the Wannier Hamiltonian through a
bond-dependent Peierls phase,
\begin{equation}
\theta_{ij}^{(\ell)}(\mathbf r_0)
=
-\frac{e}{\hbar}
\int_{\mathbf r_i}^{\mathbf r_j}
\mathbf A_{\ell}(\mathbf r-\mathbf r_0)\cdot d\mathbf r,
\label{eq:theta}
\end{equation}
which modifies the hopping according to
\begin{equation}
t_{ij}^{(\ell)}(\mathbf r_0)
=
t_{ij}^{(0)}
\exp\!\left[i\theta_{ij}^{(\ell)}(\mathbf r_0)\right].
\label{eq:peierls}
\end{equation}
Here $\mathbf r_0=(x_0,y_0)$ is the field displacement relative to the crystal; scanning it probes how the electromagnetic texture is transferred to goniopolar transport.

To identify the microscopic originTo identify the microscopic origin of the GP-window modulation, we
evaluate the energy-resolved directional transport distributions
\begin{equation}
\Sigma_{\alpha}(E)
\propto
\sum_{n\mathbf k}
v_{n\mathbf k,\alpha}^{\,2}
\delta(E-\varepsilon_{n\mathbf k}),
\qquad
\alpha=ab,c,
\label{eq:transport_distribution}
\end{equation}
with
$v_{ab}^{2}=(v_x^2+v_y^2)/2$ and $v_c^2=v_z^2$.
Within the constant-relaxation-time approximation, $S_\alpha=-(eT)^{-1}L_{1,\alpha}/L_{0,\alpha}$, where $L_{m,\alpha}$ are the usual energy moments of $\Sigma_\alpha(E)$ \cite{Madsen2018}.

This formulation directly connectsFinally, to distinguish the response common to opposite OAM windings
from the component that changes under reversal of $\ell$, we define
\begin{align}
Q_{\mathrm{even}}(|\ell|)
&=
\frac{Q(+|\ell|)+Q(-|\ell|)}{2},
\\
Q_{\mathrm{odd}}(|\ell|)
&=
\frac{Q(+|\ell|)-Q(-|\ell|)}{2}.
\label{eq:evenodd}
\end{align}
The present treatment is a prescribed static near-field Peierls model, not a time-dependent optical-excitation calculation. It therefore does not include absorption, carrier heating, dielectric screening, self-consistent propagation, or Floquet dynamics. The Peierls phase is used as the gauge-covariant lattice coupling of the specified vector potential; because $w_0=20$~\AA{} approaches microscopic length scales, the results should be interpreted as a controlled model of structured near-field coupling rather than a quantitative far-field optical prediction.
Additional calculations confirm that the principal GP-window
modulation is robust against variations of the coupling amplitude,
structured-field length scale, and longitudinal-field admixture;
the complete parameter sweeps and OAM--polarization comparison are
provided in the Supplemental Material.

Figure~\ref{fig:fig1} summarizes the equilibrium origin and coupling mechanism. The anisotropic Fermi surface produces distinct in-plane and cross-plane transport; the corresponding dark-state Seebeck curves and relativistic band/PDOS data are given in the Supplemental Material. The goniopolar windows therefore pre-exist in equilibrium: the structured field shifts their boundaries rather than creating goniopolarity. The Peierls-coupled Hamiltonian depends on two external coordinates, the field position $\mathbf r_0$ and OAM index $\ell$, whose effects are resolved below.

\begin{figure}[t]
\centering
\includegraphics[width=1\columnwidth]{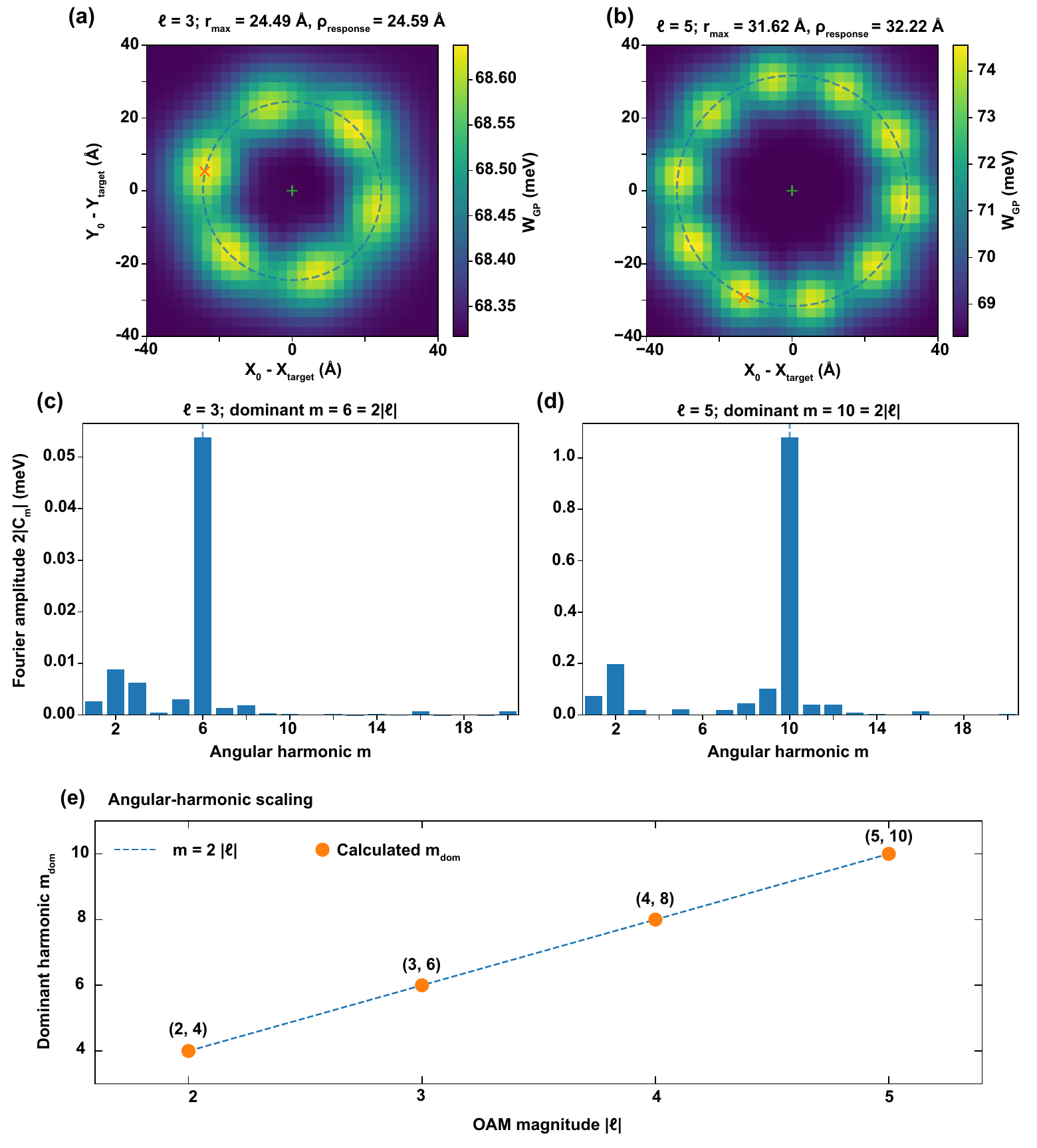}
\caption{
\textbf{Spatial imprint and angular scaling of the OAM-induced
goniopolar response in NaSn$_2$As$_2$.}
(a,b) Spatial maps of the goniopolar-window width $W_{\mathrm{GP}}$
as a function of the LG-field displacement
$(x_0-x_{\mathrm{target}},y_0-y_{\mathrm{target}})$ for
(a) $\ell=3$ and (b) $\ell=5$.
The response forms an annular pattern whose characteristic radius
closely follows the radial maximum of the $p=0$ LG envelope,
$r_{\max}=w_0\sqrt{|\ell|/2}$:
$r_{\max}=24.49$~\AA{} and
$\rho_{\mathrm{response}}=24.59$~\AA{} for $\ell=3$, and
$r_{\max}=31.62$~\AA{} and
$\rho_{\mathrm{response}}=32.22$~\AA{} for $\ell=5$.
Dashed circles indicate the corresponding LG-envelope maxima.
(c,d) Fourier decomposition of $W_{\mathrm{GP}}$ along the
response ring for (c) $\ell=3$ and (d) $\ell=5$, showing dominant
angular harmonics $m=6$ and $m=10$, respectively.
(e) Dominant angular harmonic $m_{\mathrm{dom}}$ as a function of
$|\ell|$. Independent calculations for $\ell=2$ and $\ell=4$,
combined with the $\ell=3$ and $\ell=5$ results, yield
$m_{\mathrm{dom}}=4,6,8,10$ for $|\ell|=2,3,4,5$, respectively.
Within the calculated range, the angular multiplicity therefore
follows $m_{\mathrm{dom}}=2|\ell|$.
}
\label{fig:fig2}
\end{figure}

We next examine whether the spatial structure of the applied
Laguerre--Gaussian field is transferred to the goniopolar response.
For a fixed OAM index $\ell$, we scan the beam center
$\mathbf r_0=(x_0,y_0)$ relative to the crystal and evaluate the
resulting goniopolar-window width $W_{\mathrm{GP}}(\mathbf r_0)$.
Figures~\ref{fig:fig2}(a) and \ref{fig:fig2}(b) show the resulting
two-dimensional response maps for $\ell=3$ and $\ell=5$,
respectively. In both cases, the response is strongly nonuniform and
forms an annular pattern centered on the optical-vortex core.

The radial position of this response closely follows the characteristic
radius of the $p=0$ LG envelope,
\begin{equation}
r_{\max}=w_0\sqrt{\frac{|\ell|}{2}}.
\end{equation}
For $\ell=3$, the field maximum occurs at
$r_{\max}=24.49$~\AA{}, while the calculated goniopolar response is
maximal at $\rho_{\mathrm{response}}=24.59$~\AA{}.
For $\ell=5$, these radii are $31.62$ and $32.22$~\AA{},
respectively. The outward displacement of the response maximum with
increasing $|\ell|$ therefore tracks the radial expansion of the LG
field. This correspondence provides a direct spatial link between the
imposed electromagnetic texture and the modulation of the goniopolar
window.

The annular response is not, however, azimuthally uniform. To
quantify its angular structure, we evaluate $W_{\mathrm{GP}}$ along
the response ring and decompose the angular dependence into Fourier
harmonics,
\begin{equation}
W_{\mathrm{GP}}(\phi)
=
\overline{W}_{\mathrm{GP}}
+
\sum_{m\geq1}
\left[
a_m\cos(m\phi)+b_m\sin(m\phi)
\right],
\label{eq:fourier_wgp}
\end{equation}
with $C_m=\sqrt{a_m^2+b_m^2}$.

As shown in Figs.~\ref{fig:fig2}(c) and \ref{fig:fig2}(d), the
dominant nonzero harmonic occurs at $m=6$ for $\ell=3$ and at
$m=10$ for $\ell=5$. The angular modulation therefore follows
\begin{equation}
m=2|\ell|.
\label{eq:m2l}
\end{equation}
Thus, increasing the OAM index modifies not only the radial scale of
the response but also its angular multiplicity. The observed frequency
doubling has a natural response-level interpretation. A real
winding-dependent bond perturbation contains conjugate angular
components, schematically
$\delta t_{\ell}(\phi)=a_{\ell}e^{i\ell\phi}+a_{\ell}^{*}e^{-i\ell\phi}$.
An even transport observable can therefore contain interference terms
$\delta t_{\ell}^{2}\supset
a_{\ell}^{2}e^{i2\ell\phi}+(a_{\ell}^{*})^{2}e^{-i2\ell\phi}$,
which generate $\cos(2\ell\phi)$ and hence a $2|\ell|$ harmonic.
We emphasize that this is a leading symmetry/response argument rather
than a universal theorem; the independent $|\ell|=2$--5 calculations
in Fig.~\ref{fig:fig2}(e) establish the scaling within the present
Peierls-coupled model.

Together, the radial and angular scalings show that the goniopolar response encodes the spatial degrees of freedom of the applied field.

These results establish beam position as an independent control
coordinate for goniopolarity. We next turn from this real-space
structure to the dependence on the OAM index itself and examine how
the individual goniopolar pockets evolve under reversal and increasing
magnitude of $\ell$.

\begin{figure}[t]
\centering
\includegraphics[width=1\columnwidth]{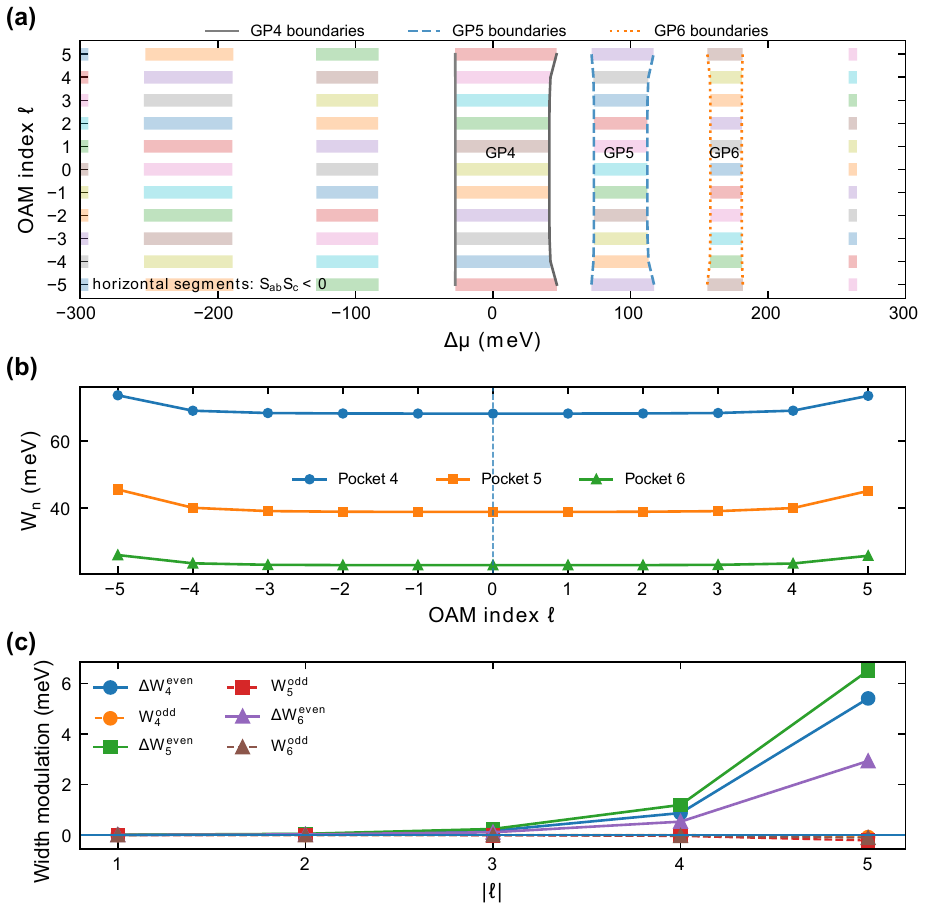}
\caption{\textbf{OAM-dependent modulation of goniopolar windows.}
(a) Goniopolar (GP) intervals at $T=300$ K as a function of chemical-potential
shift $\Delta\mu$ and OAM index $\ell$. Horizontal segments identify regions
satisfying $S_{ab}S_c<0$; the boundaries of representative pockets GP4, GP5,
and GP6 are highlighted. All three representative pockets persist throughout
the calculated range $\ell=-5,\ldots,+5$, while their boundaries shift with
$\ell$.
(b) Corresponding widths $W_n=\mu_n^{+}-\mu_n^{-}$ of GP4, GP5, and GP6.
The widths increase most strongly at large $|\ell|$, with GP5 exhibiting an
approximately 17\% enhancement at $|\ell|=5$ relative to the zero-OAM
structured-field reference.
(c) Even and odd components of the width modulation as functions of $|\ell|$.
The response common to opposite windings (even component) dominates over the
difference between $+\ell$ and $-\ell$ (odd component). Because the scan is
performed at fixed polarization $\sigma=+1$, the odd component should not be
interpreted as an isolated OAM-handedness response. The four-state
$(\ell,\sigma)$ analysis in the Supplemental Material shows that this smaller
sign-dependent correction is associated primarily with the relative
OAM--polarization configuration, conveniently labeled by
$\chi=\ell\sigma$.}
\label{fig:fig3}
\end{figure}

Having established the spatial imprint of the structured field, we now
examine how the goniopolar phase space evolves with the OAM index
$\ell$. Figure~\ref{fig:fig3}(a) summarizes the GP regions as a
function of chemical-potential shift $\Delta\mu$ and $\ell$. Each
horizontal segment denotes an interval in which
$S_{ab}S_c<0$, while its endpoints correspond to successive Seebeck
sign-change boundaries. Multiple GP pockets are present throughout the
range of OAM indices considered, demonstrating that the structured
field primarily displaces and reshapes pre-existing GP regions rather
than creating or eliminating the goniopolar state.

We track each pocket through $W_n(\ell)=\mu_n^+(\ell)-\mu_n^-(\ell)$ and its center $\mu_{n,c}=(\mu_n^++\mu_n^-)/2$, separating width modulation from displacement in chemical-potential space.

Figure~\ref{fig:fig3}(b) shows that all three representative pockets vary systematically with $|\ell|$. Most prominently, GP5 increases from approximately $38.8$~meV at $\ell=0$ to $45.5$ and $45.1$~meV for $\ell=-5$ and $+5$, an enhancement of about 17\%. GP4 and GP6 show the same qualitative trend with smaller relative changes.

Here $\ell=0$ denotes the zero-OAM structured-field reference, not
the field-free (dark) system.

An important feature of Fig.~\ref{fig:fig3}(b) is the approximate
symmetry between opposite values of $\ell$. To separate the response
associated with the magnitude of the winding from the difference
between opposite windings, we define for each GP pocket
\begin{align}
\Delta W_n^{\mathrm{even}}(|\ell|)
&=
\frac{W_n(+|\ell|)+W_n(-|\ell|)}{2}
-W_n(0),
\\
\Delta W_n^{\mathrm{odd}}(|\ell|)
&=
\frac{W_n(+|\ell|)-W_n(-|\ell|)}{2}.
\label{eq:width_evenodd}
\end{align}
Figure~\ref{fig:fig3}(c) shows that the even component dominates the
width modulation, whereas the odd component remains comparatively
small. Because these curves are evaluated at fixed polarization
($\sigma=+1$), the odd-in-$\ell$ quantity should be interpreted only
as the part that changes when the winding is reversed under that fixed
polarization condition. The principal control of the GP-window width is
therefore associated with increasing $|\ell|$, while reversal
$\ell\rightarrow-\ell$ produces a substantially weaker correction.

This hierarchy is particularly clear for GP5 at $|\ell|=5$. The two
opposite windings produce widths of approximately $45.10$ and
$45.53$~meV, corresponding to a normalized difference of less than
$1\%$. In contrast, both represent a substantial increase relative to
the $\ell=0$ reference. The response therefore cannot be described
primarily as a sign-of-$\ell$ switching effect. Instead, the dominant
behavior is an OAM-magnitude-dependent reshaping of the goniopolar
phase boundaries, accompanied by a smaller contribution that
distinguishes opposite windings at fixed polarization.

Amplitude, beam-waist, and longitudinal-component sweeps preserve this hierarchy. A four-state $(\ell,\sigma)$ test further gives $W_5(+5,+1)\simeq W_5(-5,-1)$ and $W_5(-5,+1)\simeq W_5(+5,-1)$, approximately organizing the small residual by the relative helicity $\chi=\ell\sigma$ rather than OAM sign alone; full controls are reported in the Supplemental Material.

Thus $\mathbf r_0$ controls where the response is modified, whereas $|\ell|$ controls the reconstruction in chemical-potential space. We next identify the microscopic origin of these boundary shifts.

\begin{figure}[t]
\centering
\includegraphics[width=1\columnwidth]{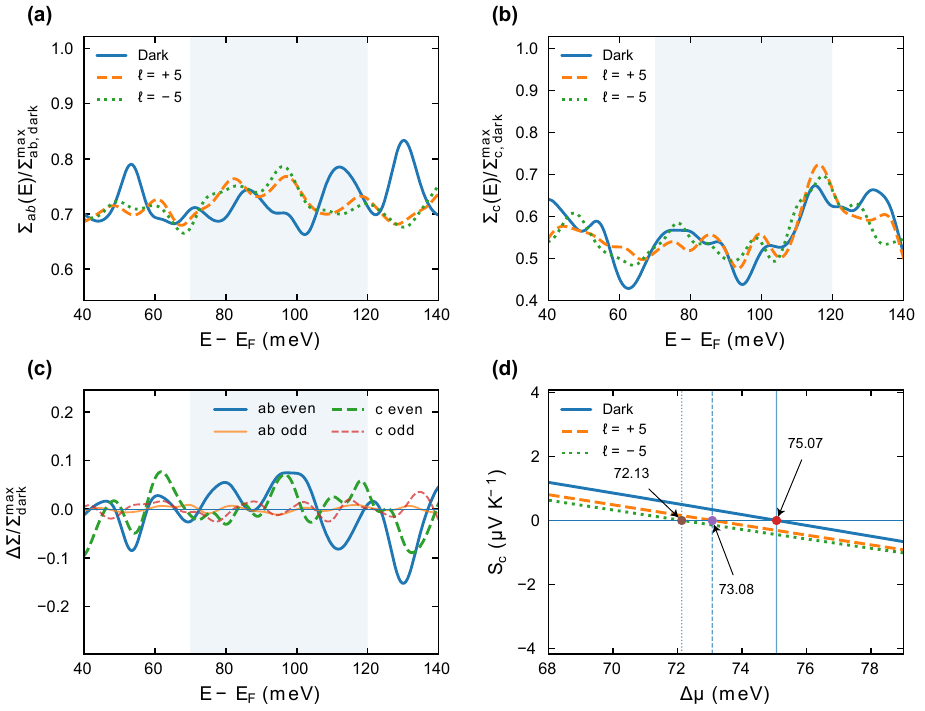}
\caption{
\textbf{Microscopic transport mechanism underlying structured-light control of goniopolarity.}
(a,b) Energy-resolved transport distributions for the dark state and $\ell=\pm5$, showing (a) the in-plane component $\Sigma_{ab}(E)$ and (b) the cross-plane component $\Sigma_c(E)$, normalized by the corresponding maximum dark-state values.
The shaded energy interval highlights the region relevant to the analyzed goniopolar boundary.
(c) Structured-field-induced changes in the transport distributions, decomposed into even- and odd-in-$\ell$ contributions for the $ab$ and $c$ directions.
The even component provides the dominant common response to $|\ell|$, while the smaller odd component measures the difference between opposite OAM windings.
(d) Cross-plane Seebeck coefficient $S_c$ near its zero crossing for the dark state and $\ell=\pm5$.
The zero crossing shifts from $\Delta\mu=75.07$~meV in the dark to $73.08$~meV for $\ell=+5$ and $72.13$~meV for $\ell=-5$.
The displacement of the Seebeck sign-change boundary connects the structured-field-induced redistribution of directional transport to the modulation of the goniopolar windows.
}
\label{fig:fig4}
\end{figure}

We finally examine the microscopic origin of the structured-field
modulation of the goniopolar phase boundaries. Within semiclassical
transport, the Seebeck coefficient is controlled not simply by the
density of electronic states, but by the energy dependence of the
directional transport distribution $\Sigma_{\alpha}(E)$ within the
thermal window around the chemical potential. The Peierls phases
introduced by the structured vector potential modify the Wannier
Hamiltonian and consequently redistribute the band velocities entering
$\Sigma_{ab}(E)$ and $\Sigma_c(E)$.

Figures~\ref{fig:fig4}(a) and \ref{fig:fig4}(b) compare the
energy-resolved in-plane and cross-plane transport distributions for
the dark system and for the two opposite OAM windings
$\ell=\pm5$. For clarity, each directional transport distribution is
normalized to the maximum of its corresponding dark-state function.
The structured field produces energy-dependent changes in both
channels rather than a uniform rescaling of the transport response.
Moreover, the modifications of $\Sigma_{ab}(E)$ and $\Sigma_c(E)$
are different, providing a microscopic route for shifting the
relative sign-change boundaries of $S_{ab}$ and $S_c$.

To distinguish the response common to opposite windings from the
difference between them at fixed polarization, Fig.~\ref{fig:fig4}(c)
shows the corresponding even and odd changes in the transport
distributions. In the energy range relevant to the GP boundary, the even redistribution dominates: for a representative 3-meV broadening, the even-to-odd RMS ratios are about 8.5 in plane and 2.3 cross plane. This microscopic hierarchy is consistent with Fig.~\ref{fig:fig3}.

The connection between this transport redistribution and the GP phase
boundary can be seen directly in the cross-plane Seebeck coefficient.
As shown in Fig.~\ref{fig:fig4}(d), the relevant zero crossing occurs
at approximately $75.07$~meV in the dark system. The zero is shifted
to approximately $73.08$~meV for $\ell=+5$ and $72.13$~meV for
$\ell=-5$. The structured field therefore produces a common
displacement of the Seebeck boundary for the two windings, together
with a smaller difference between opposite signs of $\ell$.

Writing the two field-modified zeros as $\mu_0^+$ and $\mu_0^-$, their mean displacement relative to $\mu_0^{\rm dark}$ is $\Delta\mu_{\rm even}=(\mu_0^++\mu_0^-)/2-\mu_0^{\rm dark}\simeq-2.46$~meV, whereas their separation is $\Delta\mu_\pm=\mu_0^+-\mu_0^-\simeq0.95$~meV.

The displacement common to the two windingsThe displacement common to the two windings is therefore larger than
their mutual separation, consistent with the conclusion that the
dominant response is associated with the magnitude and spatial
structure of the applied field rather than with reversal of the OAM
index alone.

The microscopic picture that emerges is consequently
\begin{equation}
\mathbf A_{\ell}(\mathbf r)
\rightarrow
\theta_{ij}^{(\ell)}
\rightarrow
\Delta\Sigma_{ab,c}(E)
\rightarrow
\Delta S_{ab,c}
\rightarrow
\Delta\mu_n^{\pm}
\rightarrow
\Delta W_n .
\label{eq:mechanism_chain}
\end{equation}
Structured-field Peierls phases redistribute the energy- and
direction-resolved electronic transport, shifting the zeros of the
directional Seebeck coefficients. Because the GP intervals are defined
by the relative signs of $S_{ab}$ and $S_c$, these shifts move their
boundaries and modify their widths. The spatial structure and OAM
dependence identified in Figs.~\ref{fig:fig2} and
\ref{fig:fig3} can therefore be traced to a microscopic redistribution
of directional electronic transport rather than to a simple uniform
change in conductivity.

We have demonstrated structured-field control of goniopolar thermoelectricity in NaSn$_2$As$_2$ through two spatial scaling laws: $r_{\rm response}\sim r_{\max}\propto\sqrt{|\ell|}$ and, for $|\ell|=2$--5, $m_{\rm dom}=2|\ell|$. Increasing $|\ell|$ simultaneously reconstructs pre-existing goniopolar windows, including a 17\% enhancement of GP5 at $|\ell|=5$, while OAM-sign reversal at fixed polarization is a much smaller effect. Energy-resolved transport establishes the microscopic sequence $\mathbf A_\ell\rightarrow\theta_{ij}^{(\ell)}\rightarrow\Delta\Sigma_{ab,c}\rightarrow\Delta S_{ab,c}\rightarrow\Delta\mu_n\rightarrow\Delta W_n$: structured-field Peierls phases redistribute directional velocities and move the Seebeck-zero boundaries defining goniopolarity. Within this static near-field model, structured electromagnetic fields therefore program both the real-space and chemical-potential boundaries separating opposite directional carrier polarities. Together with experimental demonstrations of twisted-light control of ferroic polarization textures \cite{Nazirkar2025}, the result supports a broader materials-control paradigm in which spatially structured electromagnetic fields act as external coordinates for reconfiguring distinct forms of material order, while the microscopic coupling mechanism remains material specific.

\begin{acknowledgments}
This work was supported by funding from the Douglas Jensen '59 Materials Discovery Laboratory at RPI.
\end{acknowledgments}

\section*{Data Availability}
The data and analysis scripts that support the findings of this study are available from the corresponding author upon reasonable request. A versioned repository containing processed data and analysis scripts is being prepared for public release.

\bibliography{NaSn2As2_PRL}

\end{document}